\documentclass[journal]{IEEEtran}
\usepackage{ifpdf}
\usepackage{cite}
\ifCLASSINFOpdf
   \usepackage[pdftex]{graphicx}
\else
   \usepackage[dvips]{graphicx}
\fi
\usepackage{booktabs}
\usepackage{multirow}
\usepackage{multicol}
\usepackage{algorithm}

\usepackage[cmex10]{amsmath}
\usepackage{algorithmic}
\usepackage{array}
\ifCLASSOPTIONcompsoc
  \usepackage[caption=false,font=normalsize,labelfont=sf,textfont=sf]{subfig}
\else
  \usepackage[caption=false,font=footnotesize]{subfig}
\fi

\usepackage{hyperref}

\usepackage{amsmath}
\usepackage{autobreak}
\usepackage{booktabs}
\usepackage{algorithm}
\usepackage{algorithmic}

\newtheorem{definition}{Definition}

\begin{document}
\title{Cross-chain Interaction Model In a Fully Verified Way} % TODO: replace with your title

\maketitle
\begin{abstract}
    There are different kinds of blockchains, which have been applied in various areas. Blockchains are relatively independent systems that are apt to form isolated data islands. Then cross-chain interaction is proposed to connect different blockchains. However, the current cross-chain methods do not maintain the security of the original blockchain. They either depend on a less secure third-party system or a less secure method. This makes the cross-chain interaction less secure than the original blockchains (the security downgrade issues), or the cross-chain interaction can be done even if the paired blockchain does not exist (the blockchain invisible issue). In this paper, we first propose a system interaction model and use it to analyze the possible security issues. Based on conclusions got from the proposed model, we propose the cross-chain method that verifies the data of the paired blockchain by the consensus algorithm of the paired blockchain (the CIFuV method). With this method, the cross-chain interaction can be as the same security as in the paired blockchain. At last, we evaluate the security issues during the system interaction process, and the possibility to have the CIFuV model on the public blockchains.
\end{abstract}

% TODO: replace this section with code generated by the tool at https://dl.acm.org/ccs.cfm

% \ccsdesc[300]{Security and privacy~Systems security}
% -- end of section to replace with generated code


\section{Introduction}
Blockchain technology has been used in various areas. To overcome the existing security and scalability issues in IoT, the blockchain is converged with edge computing~\cite{2020Convergence};  to solve the highly complex and costly issues in modern healthcare systems, the blockchain is used to facilitate the electronic healthcare record~\cite{2020Blockchain}; for the code plagiarism in the open-source software projects, the blockchain is used as a traceable code-copyright management system~\cite{2021A}. In blockchain-related applications, different kinds of blockchains have been used, such as Bitcoin blockchain, Ethereum blockchain, and so on.

Blockchain islands forms~\cite{kannengiesser2020bridges} if those blockchains do not interact. The blockchains are relatively separated systems, and different blockchains have their own consensus algorithm, such as PoW (proof of work), PoS (proof of stake), etc. The data structure of the blockchain, including the block and transaction, are not the same. This results that the data in one blockchain cannot be directly put to another blockchain, and it cannot use the consensus algorithm of one blockchain to validate the data from another blockchain. Generally, each system forms a relatively independent system.

Cross-chain interaction among blockchains is required to eliminate the blockchain islands~\cite{frauenthaler2020leveraging}~\cite{wang2020virtual}. In the cross-chain interaction, transactions from different blockchains are associated. Such as one transaction of blockchain $BCa$ causes or requires another transaction in blockchain $BCb$. Notice the notation of cross-chain interoperability~\cite{buterin2016chain} is also used; while in this paper, we use cross-chain interaction as it can be applied to the interaction among more general systems. Except for the connection among blockchains, cross-chain interaction also helps to improve the processing capacity of blockchains, such as the sidechain~\cite{rovzman2021scalable}. A blockchain is a decentralized system, which makes it difficult to have the same processing capacity as the centralized one. However, the parallel blockchain helps to distribute the load balance to different blockchains, which helps to improve the processing capacity.

Current cross-chain interaction includes two aspects. One is how to ensure the atomic state changes in corresponding blockchains~\cite{herlihy2018atomic}. In this way, the state changes in related blockchains should be valid or invalid together. Otherwise, the cross-chain atomicity is lost. Hash-locking~\cite{deng2018research} or two/three-phase  commitment~\cite{kan2018multiple} is used. Another aspect is the information synchronization among blockchains. Blockchain router is one of the common solution\cite{wang2017blockchain}. Considering the case when the number of interacting blockchains is large, appropriate blockchain topologies are required to facilitate the information propagation among blockchains~\cite{su2021strongly}.

\begin{figure}[htp]
    \centering
    \includegraphics[width=3in]{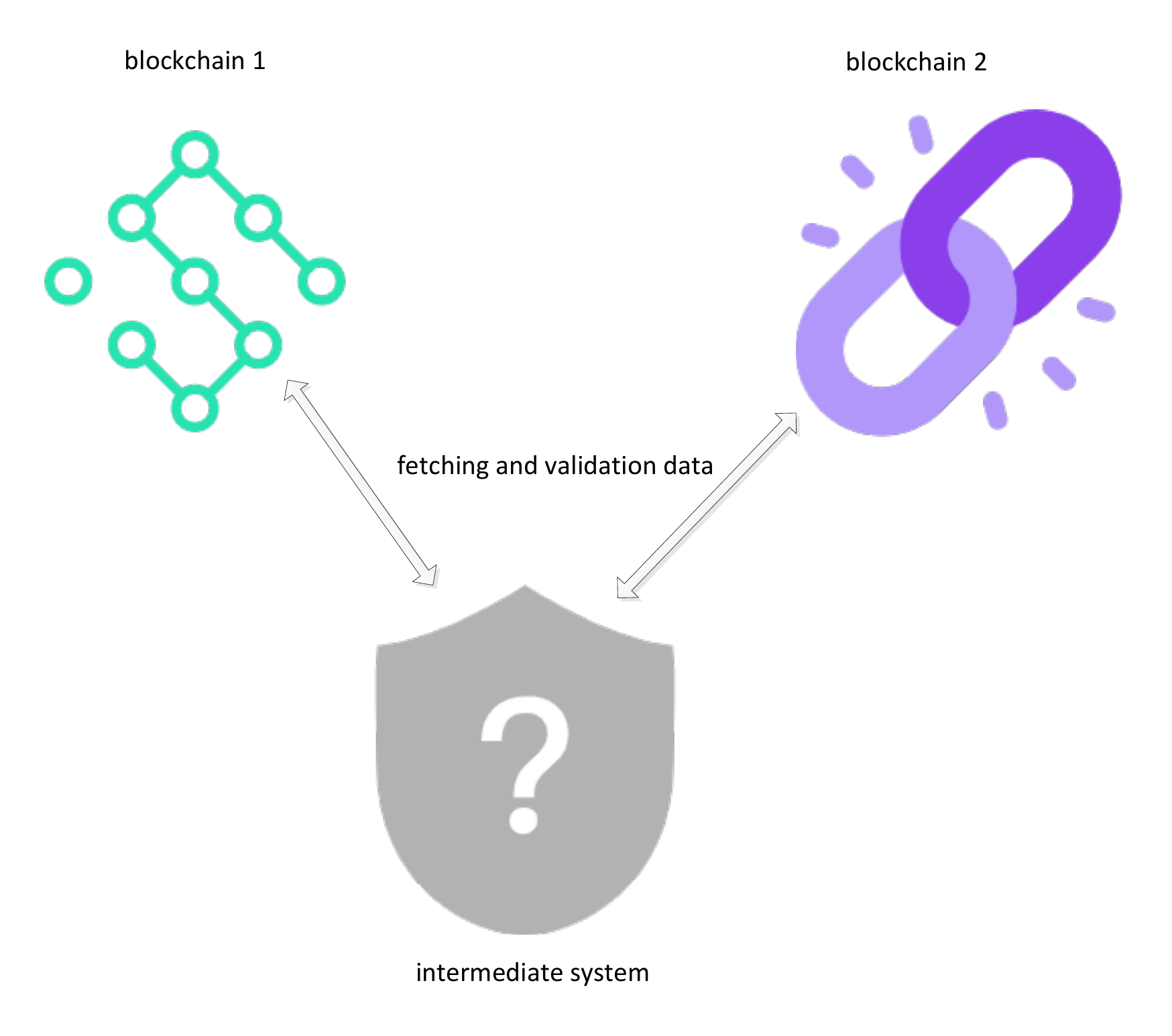}
    \caption{Cross-chain interaction by an intermediate system.}{There are two blockchains, blockchain 1 and 2, which interact through an intermediate system. Security downgrade issue may occur during the cross-chain interaction if (1) the intermediate system is less secure than blockchain 1 or blockchain 2, or (2) the method to fetch and validate the data from one blockchain (for example blockchain 1) is less secure than that blockchain (blockchain 1).}
    \label{security_downgrade_v1}
\end{figure}

However, in those cases, the interoperability among systems are done with the introduction of a less secure third-party or by a less secure method. The third-party agency is either not a blockchain system, or a blockchain system that has less secure methods. Blockchain systems are affected by those less secure factors. We call this the security downgrade issue. One example is shown as in Figure \ref{security_downgrade_v1}, in which blockchain 1 interacts with blockchain 2 by a third-party agency which is less secure than blockchain 1 and 2. As the third-party system is easier to be compromised, blockchain 1 or 2 may get wrong data more frequently. 

In this paper, we first propose the interaction model to analyze the security issues during the blockchain interaction. Then based on this model, we propose to verify the data from a blockchain (suppose blockchain $BCa$) with the consensus and corresponding data (chain of the blocks and corresponding transactions) of blockchain $BCa$. The purpose is to keep the security of the original blockchain instead of security loss (the security downgrade issue).

The major contributions of this paper are as follows. (1) We propose the data interaction model to analyze the system interaction. This model abstracts the interaction among systems as that the output data of one system is the input data of another system. This method makes the analysis of the interaction not in detail, and it helps to analyze the security issues during the interaction. (2) We point out that the security downgrade issue and the blockchain invisible issue of the data interaction model. Security downgrade issues are caused by the introduction of a less third-party or a less secure method. The blockchain invisible issue is that the cross-chain interaction is cheated by a faked (or non-existing) blockchain. (3) We propose the cross-chain interaction  (CIFuV) model with the full verification concerning the above issues. In this model, the validation of the data from a blockchain is done by the consensus of that blockchain, which achieves the same security as the original blockchain.

The rest of the paper is organized as follows. Section II describes the common interaction model and its related issues. Section III describes the proposed cross-chain interaction model with full verification (CIFuV). Section IV shows the simulation results. Section V describes the related work and section VI concludes the paper respectively.

\section{Data Interaction Model}
This section describes the proposed interaction model, and its related security issues.
\subsection{Basic Interaction Model}
The interaction among systems are in various ways: it can be that one system passes the data to another system (the data way) or one system invokes methods of another system (the method way). However, the method way can also be regarded as in the data way, as the function invocations are done by passing the corresponding parameters (data), and the execution results are returned in data format. Then the interactions can be seen as the data (the state) change of one system causes the data (the state) change in another system. Thus, we propose to model the system interaction as a data interaction model. 

Before the description of the interaction model, we describe the model of a system from the data processing point of view. We know that most systems process the data from user or other sources (such as files) as $input$ and output the result (data) as $output$. Then a system can be modeled as the data ($input$, and $output$) and the corresponding process function ($\mathbf{f_{process}}$). $output$ is generated after the processing of input by $\mathbf{f_{process}}$, as in \eqref{sys_def}. This model is called the data processing model.

\begin{equation}\label{sys_def}
    sys: output = \mathbf{f_{process}}(input)
\end{equation}
where $sys$ is the name of a system and  $\mathbf{f_{process}}$ is its processing function.

Meanwhile, there is always a validation function to check the correctness of the input data. The validation function consists the validation algorithm and the validation data. Thus, we use $\mathbf{verMethod}$ to specify the validation algorithm and use $\mathbf{verData}$ to specify the validation data. Then the data processing model can be reformed as in \eqref{sys_def_meth_data}. 

\begin{align}\label{sys_def_meth_data}
    \begin{autobreak}
sys: output = \mathbf{f_{process}}(input) 
.\mathbf{verMethod}(methodName)
.\mathbf{verData}(data)
    \end{autobreak}
\end{align}
where $methodName$ is the name of the validation method, and $data$ is the validation data.

Formula \eqref{sys_def_meth_data} can be simplified. \eqref{sys_def} is one simplified format which only reflects the relationship between the input data and the output data (without the verification function). If we only want to describe the verification part, we can use format \eqref{ver_all}. If we only want to describe the verification method, we can use format \eqref{ver_only_method}, in which the verification data is empty. If we only want to describe the verification data, we can use format \eqref{ver_only_data}, in which the verification method is empty.

\begin{equation}\label{ver_all}
sys: output = \mathbf{verMethod}(methodName).\mathbf{verData}(data)
\end{equation}

\begin{equation}\label{ver_only_method}
    sys: output = \mathbf{verMethod}(methodName).\mathbf{verData}()
\end{equation}

\begin{equation}\label{ver_only_data}
    sys: output = \mathbf{verMethod}().\mathbf{verData}(data)
\end{equation}

Based on the definition of a system, we propose the data interaction model among systems. The data interaction model is that the system interaction is through the data, in which the input of one system comes from other systems. Suppose there are two system, $sys_1$ and $sys_2$. If system $sys_1$ interacts with $sys_2$, it means that $sys_1$ gets the data from $sys_2$, as in \eqref{eq_sys_data_inter}. $sys_2$ is the \textbf{host system}, and $sys_1$ is the \textbf{guest system.} We also describe this process as data \textbf{propagates} from $sys_1$ to $sys_2$.

\begin{equation}\label{eq_sys_data_inter}
    sys_2.\mathbf{f_{process2}}(dataSys1).\mathbf{verMethod}().\mathbf{verData}()
\end{equation}
where $\mathbf{f_{process2}}$ is the processing function of $sys_2$, and $dataSys1$ is the data from $sys_1$.

If the description does not relate to the validation, \eqref{eq_sys_data_inter} can be simplified to \eqref{eq_sys_data_inter_simplified}.

\begin{equation}\label{eq_sys_data_inter_simplified}
    sys_2.\mathbf{f_{process2}}(dataSys1)
\end{equation}

If two systems interact with each other mutually as in \eqref{eq_mutual_inter}, this is a mutual interaction case. We use a keyword $\mathbf{inter}$ to specify this relationship, as in \eqref{eq_mutual_inter_key}.

\begin{equation}\label{eq_mutual_inter}
    sys_1.\mathbf{fprocess1}(dataSys2){\rm{\ and\ }}sys_2.\mathbf{fprocess2}(dataSys1)
\end{equation}

\begin{equation}\label{eq_mutual_inter_key}
    \mathbf{inter}(sys_1,sys_2)
\end{equation}

\subsection{Security Impact During the System Interaction}
In this section, we discuss the security issues during the system interaction.

\subsubsection{Quantitative Analysis}
The data interaction model is adopted to analyze the security issues during the system interaction quantitatively. Suppose the interaction is between two systems, $sys_1$ and $sys_2$, and their relationship is shown in \eqref{eq_interaction_quantitative_analysis}. If data of $sys_1$ has been changed by an attacker, the host system ($sys_2$) gets the corresponding error input ($y_1$) . 

\begin{equation}\label{eq_interaction_quantitative_analysis}
y_2 = \mathbf{f_{process2}}(y_1)
\end{equation}
where $\mathbf{f_{process2}}$ is the processing function of $sys_2$.

Generally, if $sys_1$ have a probability $p_{sys1}$ to be attacked, the host system $sys_2$ has the probability $p_{sys1}$ to get the changed data from $sys_1$.

The impact can propagate to more systems. If $sys_2$ is a guest system for other systems, this error information will propagate to other systems (host systems of $sys_2$). Suppose there is a cascaded system $CS$, as in \eqref{eq_cs_system}. Then a faked data can propagate from $sys_1$ to $sys_2$, and $sys_2$ to $sys_3$, and so on to $sys_n$.

\begin{equation}\label{eq_cs_system}
    CS: {y1=f1(x),y2=f2(y1),…,yn=fn(yn-1)}
\end{equation}

\subsubsection{Quantificational Analysis}
To further analyze the security impact, we introduce two related definitions, which are used to compare the robustness of systems quantificationally.

\begin{definition}
    \emph{broken possibility} Assume there is an attacker who tries to change data of a system (suppose system $sys_i$). If the change is successful after the attacker tries an average of $m$ times. The broken possibility, $p_{sysi}^{broken}$, is defined as $1/m$, as in \eqref{eq_broken_possibility}.
\end{definition}

\begin{equation}\label{eq_broken_possibility}
    p_{sysi}^{broken} = 1/ms
\end{equation}

\begin{definition}
    \emph{weaker system and stronger system} If one system (suppose $sys_1$) is weaker than another system (suppose $sys_2$), the broken possibility of $sys_1$ is bigger than that of $sys_2$, as \eqref{eq_weak_strong_system}. $sys_1$ is a weaker system relatively to $sys_2$, and $sys_2$ is stronger than $sys_1$.
\end{definition}

\begin{equation}\label{eq_weak_strong_system}
    p_{sys1}^{broken} > p_{sys2}^{broken}	    
\end{equation}

Notice, in some systems, there is a similar concept as the broken possibility, which can be used to compare the robustness of different systems. For example, in blockchains with the PoW consensus~\cite{nakamoto2019bitcoin}, the reciprocal of total hash calculation rate can be used as one such measurement to compare the broken possibility of PoW systems.

% With the broken possibility, the data interaction model can be reformed to \eqref{eq_system_inter_with_broken_poss}.

% \begin{equation}\label{eq_system_inter_with_broken_poss}
%     y_2 = \mathbf{f_{processing2}}(y_1, p_{sys1}^{broken})
% \end{equation}

Then, we discuss the broken possibility among several interaction systems, called the \textbf{broken possibility of the interaction systems}. The broken possibility of the interaction systems is a little different from the broken possibility of a single system, concerning whether the attacker knows which system is weak or not. If the attacker knows which system is weaker or not, he may choose to attack the weaker to save extra attempts, which is called the chosen attack model. Otherwise, he has to randomly choose one or more systems to attack or try each system in turn; in this case, all systems are treated equally, and we call this the equal attack model.

The following describes the broken possibility of the interaction systems of the equal attack model and the chosen attack model. Suppose (1) there are $n$ systems, from $sys_1$ to $sys_n$, which interact mutually, (2) an attacker attempts to change data used for interaction in those systems for its own benefit, and (3) the broken possibility of $sys_i$ is $p_{sysi}^{broken}$. 

\textbf{Equal attack model} As the attacker does not know which system is weak, he has to choose the target system(s) randomly. Then each system has an equal chance to be attacked. Thus, the chosen possibility for each system is $1/n$. The chosen possibility is the possibility that a system is chosen to attack in one attempt. This kind of attack is called the equal attack model.

With the above assumptions, the broken possibility of the whole system, $p_{all}^{broken}$, is as in \eqref{eq_equal_attack_possibility}.

\begin{equation}\label{eq_equal_attack_possibility}
    p_{all}^{broken} = max(p_{sysi}^{broken})/n 
\end{equation}
where $sys_i$ stands for any system in the interaction.

Here we give a brief proof for \eqref{eq_equal_attack_possibility}. When a system is broken, the interaction is trustless, as other systems get information mutually. Then when the first system is compromised, the interaction can be regarded as compromised. We need to find which system is the first one to be compromised. Averagely, when the number of real attempts is more than the number of the required attempt (1/$p^{broken}$), a system is broken. We notate the attempts which have been already carried out as $ra$, and the number of attempts that are further needed to break the system as $la$. Then the relationship among $la$, $ra$, and $p^{broken}$ is as in \eqref{eq_ra_la}, which is also called the \textbf{attack equation}.

\begin{equation}\label{eq_ra_la}
    \left\{ \begin{array}{l}
        la_1 = ra/n - 1/p{_{sys1}^{broken}}\\
        la_2 = ra/n - 1/p{_{sys2}^{broken}}\\
        ...\\
        la_i = ra/n - 1/p{_{sysi}^{broken}}\\
        ...\\
        la_n = ra/n - 1/p{_{sysn}^{broken}}
        \end{array} \right.
\end{equation}

When $la_i$ is 0, $sys_i$ is broken. The first compromised systems is the system of which the value $n/p{_{sysi}^{broken}}$ is the minimum. The average attack possibility is as in \eqref{eq_p_all_equal}.

\begin{equation}\label{eq_p_all_equal}
    p_{all}^{broken} = 1/min(n/p{_{sysi}^{broken}})=max(p{_{sysi}^{broken}}/n)
\end{equation}

\textbf{Chosen attack model} When an attacker knows which systems are weaker, it may choose to attack one or several systems more frequently. We use the chosen possibility, $p_{sysi}^{select}$, as the possibility to choose $sys_i$ to attack. $p_{sysi}^{select}$ matches conditions as \eqref{eq_def_select}.

\begin{equation}\label{eq_def_select}
    \left\{ \begin{array}{l}
        0 =  < p_{sysi}^{select} =  < 1\\
        \sum\limits_{i = 1}^n {p_{sysi}^{select} = 1} 
        \end{array} \right.
\end{equation}

Especially, if an attacker finds system $sys_k$ is much weaker than other systems, he can choose to attack this system only, the chosen possibility is as in \eqref{eq_attack_one_sys}.

\begin{equation}\label{eq_attack_one_sys}
    \left\{ \begin{array}{l}
        p_{sysk}^{select} = 1\\
        p_{sysn}^{select} = 0,{\rm{\ when\ }}n{\rm{! = }}k
        \end{array} \right.
\end{equation}

For the chosen attack model, the attack equation is as in \eqref{eq_chosen_attack_model}.

 \begin{equation}\label{eq_chosen_attack_model}
    \left\{ \begin{array}{l}
        la1 = ra*p{_{sys1}^{select}} - 1/p{_{sys1}^{broken}}\\
        la2 = ra*p{_{sys2}^{select}} - 1/p{_{sys2}^{broken}}\\
        ...\\
        lan = ra*p{_{sysn}^{select}} - 1/p{_{sysn}^{broken}}
        \end{array} \right.
 \end{equation}

Then we get the broken possibility of the whole system as in \eqref{eq_p_all_chosen}.

\begin{equation}\label{eq_p_all_chosen}
    p_{all}^{broken} = 1/min(p_{sysi}^{select}/p{_{sysi}^{broken}})=max(p{_{sysi}^{broken}}/p_{sysi}^{select})
\end{equation}

\eqref{eq_p_all_equal} is the special case of the chosen attack model when an interaction systems match conditions as  in \eqref{eq_chosen_to_equal}.

\begin{equation}\label{eq_chosen_to_equal}
    p_{sysi}^{broken} = 1/n
\end{equation}
where $sys_i$ stands for any system in an interaction. 

\subsubsection{Security Downgrade and System Invisible Issues}
In an interaction, if related systems have different broken possibilities, it may cause the security downgrade issue. If there is a third-party to pass the data among two systems, it may cause the system invisible issues.

\textbf{Security downgrade} Security downgrade issue is that a less secure system causes the broken possibility of the whole systems ($p_{all}^{attatck}$) is bigger than that of one or more interacting systems. Suppose there are $n$ interacting systems, from $sys_1$ to $sys_n$. If \eqref{eq_secueity_downgrade_con} matches, it causes that $p_{all}^{attatck}$ is bigger than that of $sys_j$ as in \eqref{eq_secueity_downgrade}, and we call that $sys_j$ is security downgraded by $sys_i$.
 
\begin{eqnarray}\label{eq_secueity_downgrade_con}
    p_{sysi}^{broken}/p_{sysi}^{select} > p_{sysj}^{broken}
\end{eqnarray}

\begin{equation}\label{eq_secueity_downgrade}
    p_{all}^{broken} >= p_{sysi}^{broken}/p_{sysi}^{select} > p_{sysj}^{broken}
\end{equation}

The reason is that the weaker system has more possibility (higher broken possibility) to be attacked, and then the stronger system has more possibility to get the changed data from it.

We use the interaction between two systems to explain. Assume those systems are $sys_1$ and $sys_2$, and $sys_1$ is the weaker system. In the equal attack model, if $p_{sys1}^{broken}$ is more than twice $p_{sys2}^{broken}$, $sys_2$ is security downgraded by $sys_1$, as in \eqref{eq_sd_two_eq}. In the chosen attack model, $sys_2$ is downgrade by $sys_1$ even if $p_{sys1}^{broken}$ is only one time bigger than $p_{sys2}^{broken}$, as in \eqref{eq_sd_two_ch}.

\begin{equation}\label{eq_sd_two_eq}
p_{sys1}^{broken} > p_{sys2}^{broken} * 2	
\end{equation}

\begin{equation}\label{eq_sd_two_ch}
p_{sys1}^{broken} > p_{sys2}^{broken}	, if p_{sys1}^{select}=1	
\end{equation}

\textbf{System Invisible} For two systems ($sys_1$ with $p_{sys1}^{broken}$, $sys_2$ with $p_{sys2}^{broken}$), and a third-party agent $sys_t$. We say that system $sys_1$ is invisible for $sys_2$ when the following two conditions match.

(1) $sys_2$ gets the data of $sys_1$ only from $sys_t$.

(2) $sys_2$ does not check the correctness of the data of $sys_1$ by the validation algorithm of $sys_1$, as shown in \eqref{eq_sys_invis_con}. Normally, if $sys_2$ cannot directly get data from $sys_1$, $sys_2$ should use the interface of $sys_1$ to validate whether the data is from $sys_1$ or not.

\begin{equation}\label{eq_sys_invis_con}
    sys_2: y2 = verMethod(methodNotBySys1).verData(data_{syst}^{sys_1})
\end{equation}
where $data_{syst}^{sys_1}$ is the data that gets from the third-party $sys_t$ which declares that this data is of $sys_1$, and $methodNotBySys1$ is that the validation is done by other system instead of $sys_1$.

When the above two conditions match, it indicates that $sys_1$ does not need to really exist for $sys_2$, as $sys_t$ can fake the data and there is no method to validate the correctness from the original system ($sys_1$). This results that $sys_2$ still seems to interact with $sys_1$ even if $sys_t$ fakes the data. This phenomenon is called the system invisible.

System invisible may cause the following issues.

Issues 1, the third-party may change the data due to some purpose. As the target system is invisible, it is convenient for the third-party to change the data.

Issues 2, when the third-party is weak, it may introduce a security downgrade issue.

Then there are some rules to choose a third-party to avoid the above possible security issues. If a system matches the following two rules, it is called a secure third-party.

\textbf{Security third-party rule 1} The logic of a third-party system should be publicly available and verifiable. If its logic cannot be verifiable, the third-party may change the data. If the logic is not publicly available, other systems may not have the chance to verify the logic.

\textbf{Security third-party rule 2} A third-party should be a secure enough system. Then a system with low broken possibility should be chosen.

We take the hash-locking method as an example to analyze. It is a method to keep atomicity among different blockchains. The data to unlock an asset is from another blockchain, which is passed by a user. There is no direct communication between interacting blockchains. The user acts as a third-party system, which cannot be verified. This makes users have the possibility to cheat the host blockchains that a faked blockchain exists.

\section{Cross-chain Interaction}
This section describes the cross-chain model in the full verification way. We first describe the elements to validate the data correctness of a blockchain, which can be described by a quintuple, as in (0). This quintuple is also called a $blockchain\ quintuple$ ($BCQ$).

\begin{equation}
    BCQid = (id,consensus,blocks,transactions)
\end{equation}
where $id$ is the identification of a blockchain, which is often associated with the genesis block (and in this paper, we assume that $id$ is in the genesis block), $consensus$ is the blockchain consensus algorithm, $blocks$ are the chain of blocks, and $transactions$ are the corresponding transactions.

With the blockchain quintuple, a validator can run the consensus algorithm to verify the correctness of a transaction or a block in that blockchain.

\subsubsection{Current Cross-chain method and Related Security Issues}
Currently available cross-chain methods include hash-locking method~\cite{deng2018research}, two/three-phase commit~\cite{kan2018multiple}, and blockchain router~\cite{wang2017blockchain}. There are two issues with those methods. (CI\_1) Only partial information (such as only corresponding transactions instead of the total chain of blocks ) is used in the interaction. (CI\_2) Some methods (such as the hash-locking method) rely on participants to exchange information among blockchains. 

We focus on the analysis of CI\_1, as the security issue of CI\_2 has been discussed in previous sections. For CI\_1, it may cause the security downgrade issue and the system invisible issue when partial information of the blockchain quintuple is used. (1) Lack of $id$ may cause that the data is from a different blockchain. As $id$ is related to the genesis block, this indicates that the genesis block is invisible. If the genesis is not verified, it can be the data from another blockchain. This causes the system invisible issue. (2) Lack of $consensus$ and $blocks$ indicates that the host blockchain does not check the data by the validation method of the guest blockchain. If both the consensus algorithm and the whole chain of blocks are missing, other methods are difficult to validate the data with the same security as a blockchain. Then it causes the security downgrade issue.

From the above analysis, we can see that there are different degrees of security loss in the current cross-chain interaction methods. They can be regarded as a simplified cross-chain interaction method. In this paper, we propose the cross-chain interaction model in the full verification way, which is also called the CIFuV model.

\subsubsection{Cross-chain Interaction With Full Verification (CIFuV) model}

\begin{figure}[htp]
    \centering
    \includegraphics[width=3in]{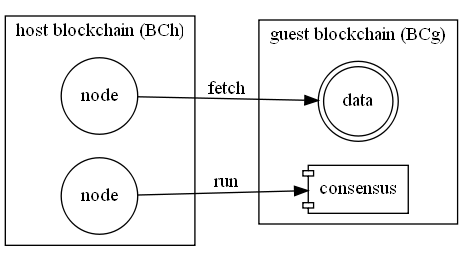}
    \caption{Cross-chain interaction with full verification.}{It has two steps: (1) the host blockchain ($BCh$) fetches data of its guest blockchain ($BCg$), and (2) $BCh$ uses the consensus algorithm of $BCg$ to verify the validation of the data from $BCg$.}
    \label{interaction_model}
\end{figure}

The cross-chain interaction with full verification (CIFuV) model is a cross-chain interaction model, in which one blockchain (the host blockchain) verifies the data from another blockchain (the guest blockchain) by (1) the consensus algorithm of the guest blockchain, and (2) the chain of blocks and other two elements (in the $blockchain\ quintuple$) of the guest blockchain. This kind of verification is called cross-chain full verification. Figure \ref{interaction_model} shows this process.

The elements in the $blockchain\ quintuple$ can be divided into two types. The consensus is the method of the verification and the other three elements can be seen as the data of the verification. Then the full verification model can be divided into two aspects, the consensus verification, and the necessary data verification. 

\subsubsection{Cross-chain Consensus Verification}
The host blockchain runs the consensus algorithm of the guest blockchain to verify whether data from the guest blockchain is correct or not. The nodes of the host blockchain can be seen as the verification nodes of the guest blockchain. This kind of consensus is called the cross-chain consensus, as it aims to verify data for the guest blockchain instead of the host blockchain. 

However, there is no need for nodes of the host blockchain to contribute to the guest blockchain as mining is not required by the validation. The nodes of the host blockchain run a subset of the consensus of the guest blockchain. 

Then the consensus algorithm of the host blockchain has two parts, one is the consensus algorithm of the host blockchain and another is the verification of the guest blockchain, as in \eqref{eq_con_set}.

\begin{equation}\label{eq_con_set}
    \mathbf{con_{h}}={con.own_{h}, con.ver_{g}}
\end{equation}
where $con.own_{h}$ is the consensus algorithm of the host blockchain, $con.ver_{g}$ is the verification part of the guest blockchain, and $\mathbf{con_{h}}$ is all the consensus algorithms on the host blockchain, which can be seen as the set of consensus algorithm run on the host blockchain.

\subsubsection{Necessary Cross-chain Data Validation}
The data for the cross-chain verification includes three elements from the guest blockchain, $id$, $blocks$, and $transactions$. Those data is called the cross-chain data for the host blockchain. (1) The chain of blocks ($blocks$) is required for the cross-chain verification. (2) $id$ can be got from the genesis block, which is in blocks. (3) For the required transactions ($transactions$), there are two cases. (a) In most blockchains, there is a Merkel tree of the transactions, which can be used to verify the validation of a transaction with related transactions along the Merkel tree. (b) For the system without the Merkel root mechanism, all transactions are required. Then we describe the data for the validation as the necessary data. The necessary data includes the chain of the blocks and corresponding transactions (as that in case a or case b depending on the blockchain types).

With the consensus validation and the necessary data validation, the validation of the cross-chain data is based on the $blockchain\ quintuple$ of the guest blockchain, as in \eqref{eq_ver_guest}.

\begin{equation}\label{eq_ver_guest}
ver= verMethod(consensus_{g}).verData(ci_{g}, blocks_{g}, transactions_{g})
\end{equation}
where $consensus_{g}$ is the consensus algorithm of the guest blockchain, $ci_{g}$ is the guest blockchain's identifier, $blocks_{g}$ is the chain of blocks of the guest blockchain, and $transactions_{g}$ is the necessary transactions of the guest blockchain. 

\subsubsection{Relationship Between Host blockchain and Guest blockchain}
\begin{figure}[htp]
    \centering
    \includegraphics[width=3in]{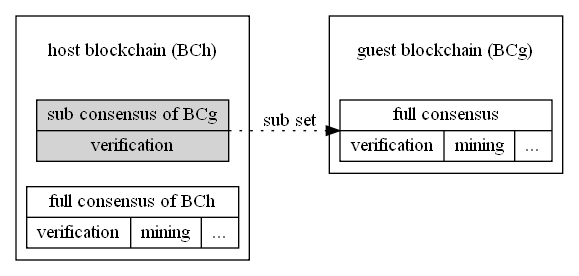}
    \caption{Relationship between the host blockchain ($BCh$) and the guest blockchain ($BCg$).}{$BCh$ contains a sub set of $BCg$’s functions. For example, $BCh$ only runs the verification part instead of the whole consensus (including the verification and mining part) of $BCg$.}
    \label{interaction_model_subset}
\end{figure}

This section describes the relationship between the host blockchain and the guest blockchain. The host blockchain contains a “sub” set of functions of the guest blockchains, which is shown in Figure \ref{interaction_model_subset}. We summarize them as follows.

(1) The consensus of the host blockchain is a sub-set of the guest blockchain. It only contains the part to validate the data of the guest blockchain (no mining functions to save consumption of CPU resources).

(2) The blockchain data is also a sub-set of the guest blockchain. The chain of blocks is required, while not all transactions are required when the Merkel tree is used.

(3) The nodes of the host blockchain are a part of the P2P network of the guest blockchain, which aims to fetch corresponding data from the guest blockchain.

\subsection{Characteristics of CIFuV model}
The following characteristics are drawn from the CIFuV model, which is the requirement for blockchains systems to interact.

(1) Asynchronous process

The verification and mining processes for two blockchains are asynchronous processes. Different blockchains have different mining periods. It is difficult to synchronize mining periods among different blockchains.

As the process among systems is asynchronous, it is required to have the mechanism to wait or callback to wait for visible issues in another blockchain.

(2) Synchronization-Keeping.

Currently, it takes a relatively long time to synchronize the blockchain data for the first time (called the first-time synchronization), as all previous blocks and related transactions will be synchronized. For example, it takes several days to synchronize the data of Bitcoin at the first time (the initial block download)\footnote{https://bitcoin.org/en/full-node, visit on 2021-04-09}. Although, the synchronization speed will improve in the future network, such as 6G; while currently it needs a long time. Thus, we propose the synchronization-keeping technology. The synchronization-keeping is to keep the synchronization with the guest blockchain after the first-time synchronization. The later synchronization only needs to get the newly-generated data, which is an increment synchronization.

In this way, when two blockchains want to interact, they first connect to each other and synchronize the data that already exists (the first-time synchronization). When the first-time synchronization completes, it continues to synchronize the incremental blocks and transactions.

(3) Double rebranch issues. In the cross-chain interaction, there are at least two blockchains, the host blockchain, and its guest blockchain. Then the rebranch can happen in the host blockchain and the guest blockchain, this is called the double rebranch issue.

For one blockchain, it is only required to wait for enough number of blocks in its own blocks. In the cross-chain scenario, it is required to wait for enough number of blocks in both the host blockchain and the guest blockchain.

\section{Evaluation}
The evaluations are done in two aspects. (1) When one system is weaker than others, there may be security downgrade issues. This is the first evaluation, security impact during the system interaction. (2) In the CIFuV model, it is required to run the validation part of the consensus algorithm of the guest blockchain. The validation nodes of the host blockchain need to run two consensus algorithms (at least the validation parts). We elevate the possibility to run two consensus algorithms (of two public blockchains) on one node.

\begin{figure*}
    \centering
    \includegraphics[width=8in]{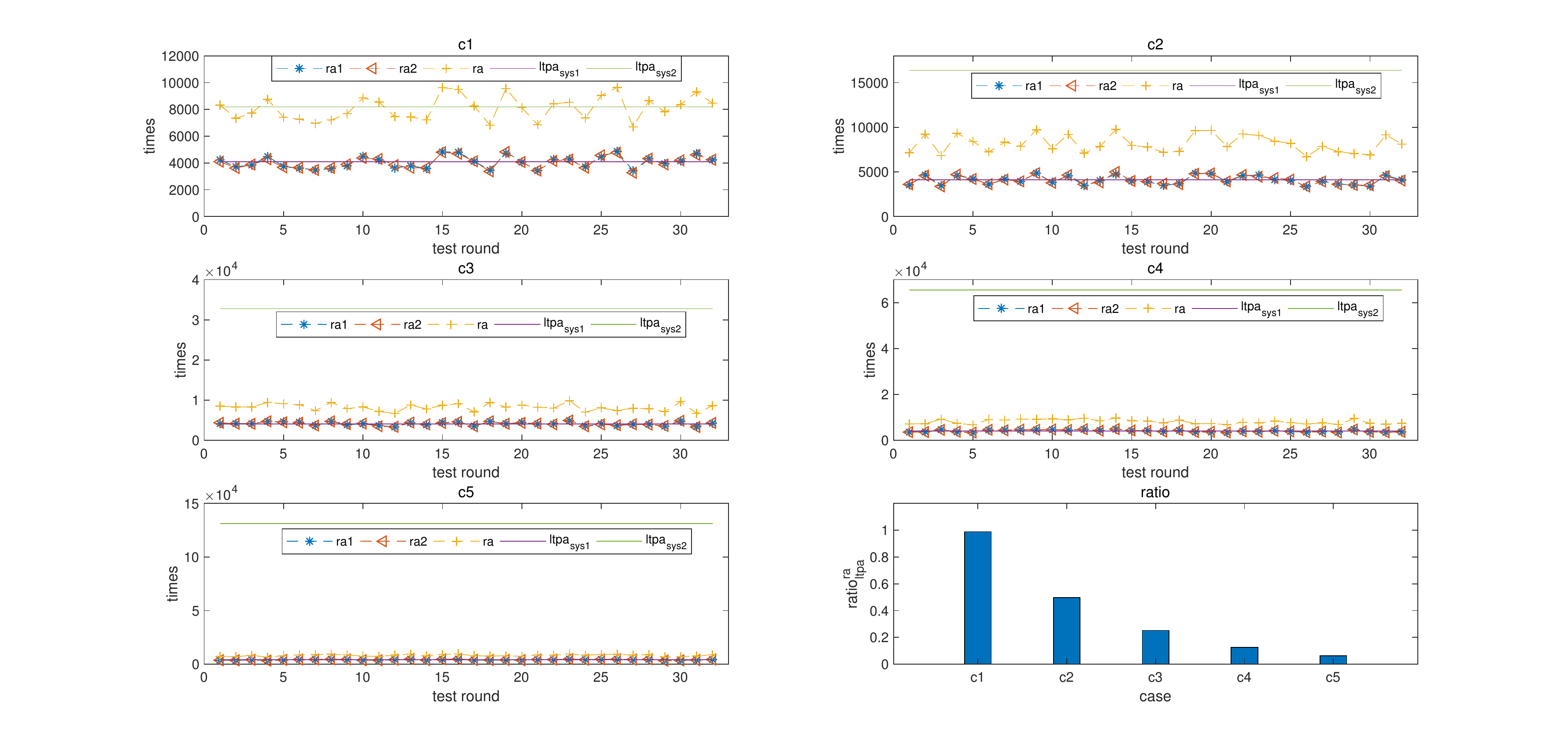}
    \caption{The number of the real attack ($ra$) for a successful attack with equal broken possibility for two systems.}{From the top to down, the first five parts are cases for c1, c2, c3, c4, and c5 in table \ref{t_number_of_ltpa}. As the number of the one-time predicted attack ($otpa$) is the same as $ra$ when success, it overlaps with the curve of $ra$. Then $otpa$ is not shown in this figure. The last ('ratio') part is the tendency of $ratio^{ra}_{ltpa}$ as defined in \eqref{eq_ratio_ra_ltpa}.}
    \label{Equal_chance_attack}
\end{figure*}

\begin{figure*}
    \centering
    \includegraphics[width=8in]{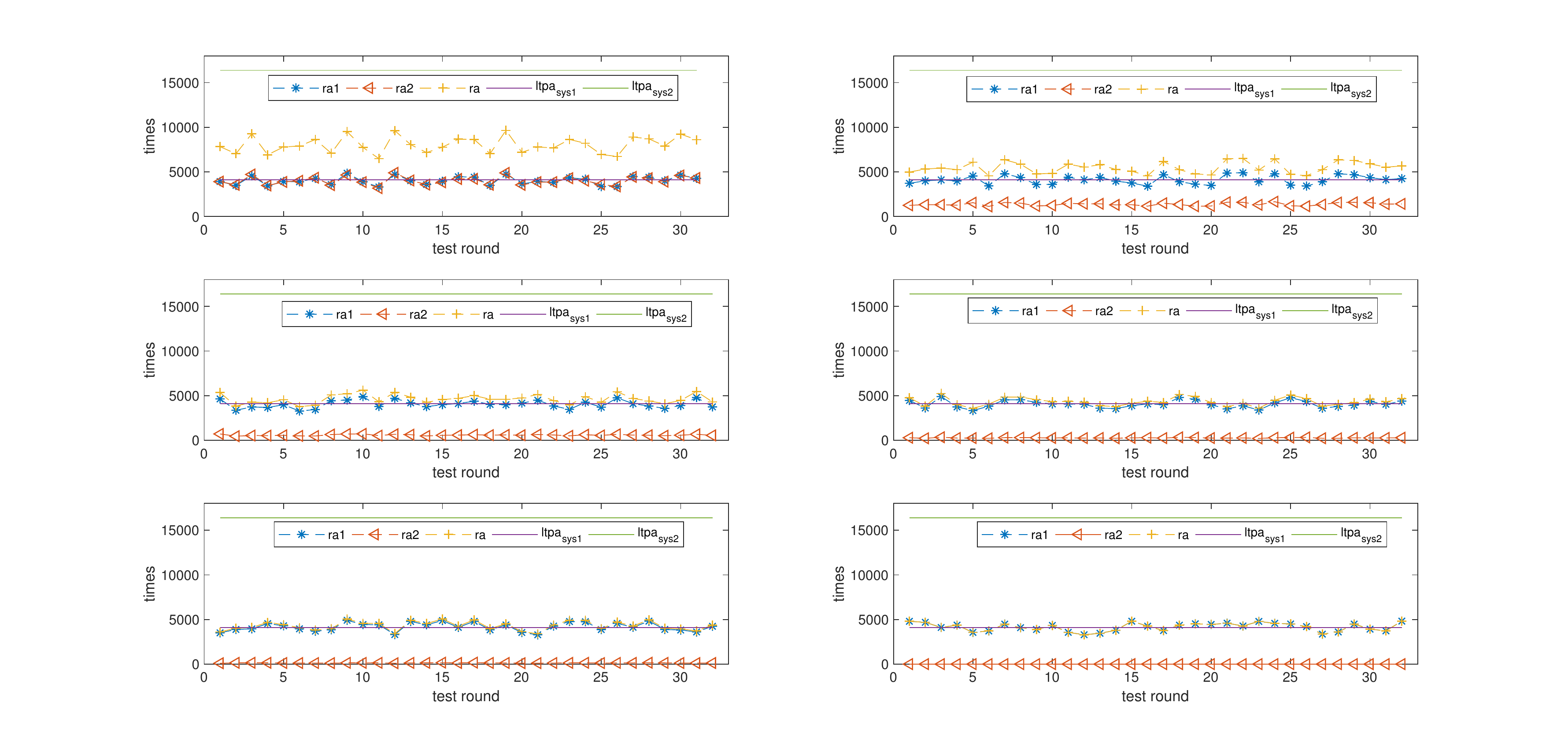}
    \caption{The results of $ra$ with different choose possibility for the weaker systems (1/2, 3/4, 7/8, 15/16, and 1)}
    \label{Nonequal_chance_attack}
\end{figure*}

\begin{figure*}
    % \centering
    \includegraphics[width=8in]{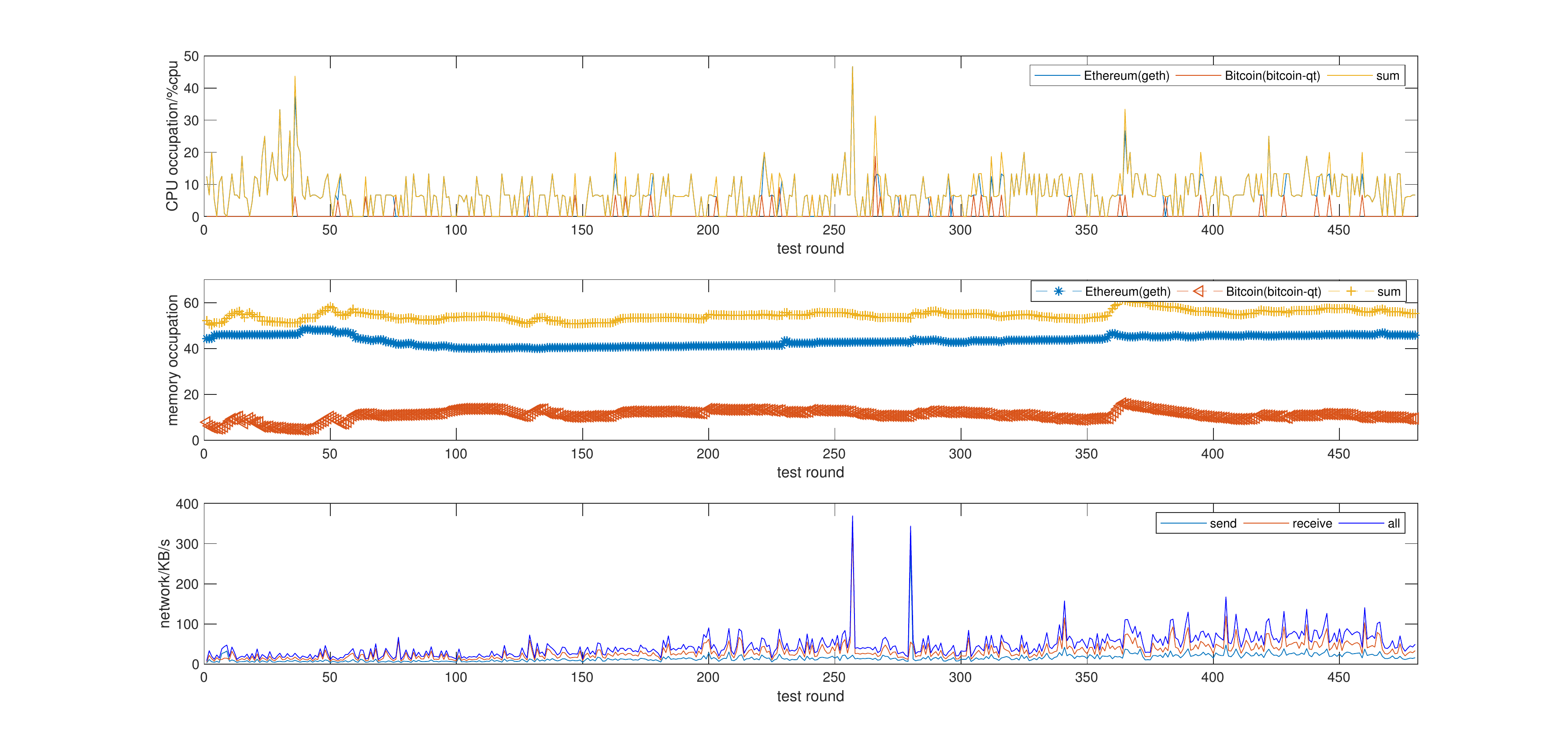}
    \caption{Consumption of CPU, memory, and network bandwidth}
    \label{cpu_memory_network}
\end{figure*}

\subsection{Security Impact During the System Interaction}
\textbf{Measurement in the evaluation} The reciprocal of the broken possibility ($p^{broken}$) is the average number of the attempts to have a successful break. It is not the number of attempts in one specific time, as it may vary in a certain range due to the current status of that system. Then the reciprocal of the broken possibility is called the number of the long-term predicted attempt (or abbreviated as $ltpa$). And the number of attempts required by a system in a specific time to have a successive break is called the number of the one-time predicted attempt (or abbreviated as $otpa$). It requires $ra$ (the attempts which have been already carried out, the real attack) is no less than $otpa$ for a successful break, as in \eqref{eq_rel_ra_otpa}.

\begin{equation}\label{eq_rel_ra_otpa}
    ra > otpa
\end{equation}

\textbf{Simulation of interacting systems} There are two simulated systems in the evaluation, $sys_1$ and $sys_2$. Those systems are simulated ones, as the broken possibility may change greatly in real systems. 

$Sys_1$ is weaker than $sys_2$, with $ltpa$ ($ltpa_{sys1}$) as 4096. $Sys_2$ is stronger, of which $ltpa$ ($ltpa_{sys2}$) is $n$ ($n$ > 1) times of that of $sys_1$, as in table \ref{t_number_of_ltpa}. To simulate the number of the one-time predicted attempt ($otpa$), we randomly generate it within the range as shown in \eqref{eq_otpa_rel}.

\begin{equation}\label{eq_otpa_rel}
    0.8 ltpa =< otpa =< 1.2 ltpa    
\end{equation}

\begin{table}[!t]
    \caption{Number of long-term predicted attempt}\label{t_number_of_ltpa}
     \centering
     \begin{tabular} {cccccc}
      \toprule
      Case & c1 & c2 & c3 & c4 & c5 \\
      \midrule
      Sys1 & 4096 & 4096 & 4096 & 4096 & 4096\\
      Sys2 & 8192 & 16384 & 32768 & 65536 & 131072\\
      n & 2 & 4 & 8 & 16 & 32\\
      \bottomrule
    \end{tabular}
   \end{table}

The simulation of the attack model. An attacker randomly chooses to attack one of the two systems in each attempt. We use $ra1$ to record the number of real attack for $sys_1$ and $ra2$ for that of $sys_2$. For any system, when the number of the test attempt is equal to or more than the one-time predicted attempt, it is regarded as a successful attack, as in \eqref{eq_ra_otpa_sys}. 

\begin{equation}\label{eq_ra_otpa_sys}
    ra1 > otpa_{sys1}{\rm{\ or\ }}ra2 > otpa_{sys2}
\end{equation}
		
The total number of the attempt for the attacker is the sum of the attempts for both systems, as in \eqref{eq_all_ra_and_sub_ra}.

\begin{equation}\label{eq_all_ra_and_sub_ra}
    ra = ra1 + ra2    
\end{equation}

\subsubsection{Equal Attack Case}
In the equal attack case, the attacker has the equal chance to attack $sys_1$ or $sys_2$. The simulation of the scenario is that one number (either 1 or 2) is generated by the random method of Perl from its “Math::Random” module \footnote{https://metacpan.org/pod/Math::Random}. If the number is 1, the attack is for $sys_1$ ($ra1$ increases 1), and if the number is 2, the attack is for $sys2$ ($ra2$ increases 1). The pseudocode is shown in algorithm \ref{alg_random_system_choice}, in which line \ref{alg_lb_gn_random} generates the random number 1 or 2. Lines from \ref{alg_random_system_choice_choose_start} to \ref{alg_random_system_choice_choose_end} are to choose a system. The evaluation results are shown in Figure \ref{Equal_chance_attack}.

\begin{algorithm}[htb]
    \caption{ Random system choice }
    \label{alg_random_system_choice}
    \begin{algorithmic}[1] 
        \STATE {r =generateRandom(1, 2)} \label{alg_lb_gn_random}
        \IF {1 == r} \label{alg_random_system_choice_choose_start}
            \STATE { ra1 += 1 }
        \ENDIF
        \IF {2 == r}
            \STATE { ra2 += 1 }
        \ENDIF \label{alg_random_system_choice_choose_end}
    \end{algorithmic}
  \end{algorithm}

There are two types in Figure \ref{Equal_chance_attack}, concerning whether there are security downgrade issues for the more secure system ($sys_2$) or not, type 1 (at1) with security downgrade issues, and type 2 (at2) without security downgrade issues.

In the first type (at1), $sys_2$ has been security downgraded (in c2, c3, and c5 cases) as $ra$ is lower than $ltpa_{sys2}$. This happens when $ltpa$ of $sys_2$ is more than 2 times of $lpta$ of $sys_1$. Takes part 'c5' in Figure \ref{Equal_chance_attack} as example, the difference of $ltpa$ between $sys_2$ and $sys_1$ is 32; $ltpa_{sys2}$ is 131072, while the maximum of $ra$ is only 9509, which is less than 1/10 of $ltpa_{sys2}$.

Another interesting point is that the difference between $ra$  and $ltpa_{sys2}$ increases when the secure difference between $sys_1$ and $sys_2$ increases. The tendency can be seen from the first part to the fifth part in Figure \ref{Equal_chance_attack}. This indicates that a less secure system makes the whole interaction less secure no matter how secure other systems are.

To have a more clear relationship between $ra$ and $ltpa_{sys2}$, we adopt the ratio of $ra$ to $ltpa_{sys2}$, as in \eqref{eq_ratio_ra_ltpa}. The average value of $ra$ is used, as it changes in each test round. The corresponding results are shown in the last ('ratio') part of Figure \ref{Equal_chance_attack}.

\begin{equation}\label{eq_ratio_ra_ltpa}
    ratio^{ra}_{ltpa} = (\sum\limits_1^n {ra/n)/ltpa_{sys2}}
\end{equation}

For the second case (at2), $ra$ changes around $ltpa_{sys2}$ as in c1 (the first part in Figure \ref{Equal_chance_attack}). In some test round, $ra$ is higher than $ltpa$. Such as in test round 4, the attacker has tried 8751 times, which is bigger than $ltpa_{sys2}$ (8192). Totally there are 15 times (in all 31 test rounds) in which $ra$ is higher than $ltpa_{sys2}$. However, in the left 16 test rounds, $ra$ is lower than $ltpa$. Such as in test round 3, the attacker has tried 7732 times, which is lower than $ltpa_{sys2}$. 

The reason, for which $ra$ is near $ltpa_{sys2}$ in c1 case, is that the attacker does not know which system is more secure, it has to try each system with an equal probability. Then averagely a system only gets one attempt of two real attacks. And as $ltpa$ of $sys_2$ (the more secure system) is twice as that of $sys_1$ (the less secure system), $ra$ is near to $ltpa$ of $sys_2$ in average. 

The condition of at2 can be regarded as a demarcation point for the equal attack between two systems. When the difference of $lpta$ is bigger than 2 times, it will cause a security downgrade issue; otherwise, it will not cause the security downgrade issue.	

\subsubsection{Chosen Attack Case}
The above is the result when an attacker does not know which system is weaker. If the attacker knows which system is weaker, it can adopt a more optimized way to attack the weaker system. Thus, we verify the scenario when the attacker knows which system is weaker and has a different chosen possibility. 

The verification case is c2 case in table \ref{t_number_of_ltpa}, in which  $lpta_{sys2}$ is 4 times of $lpta_{sys1}$. Choose possibilities for the weaker system ($sys_1$) are 1/2, 3/4, 7/8, 15/16, 31/32, and 1 separately. The results are shown in Figure \ref{Nonequal_chance_attack}.

From Figure \ref{Nonequal_chance_attack}, we see that all curves of $ra$ are under the curves of $lpta_{sys2}$. In the first diagram, the minimum value of $ra$ is 9652, which is lower than the value of $lpta_{sys2}$, 16384. When the chosen possibility increases from 1/2 to 1, the curve is farther from that of $lpta_{sys2}$ and closer to that of $lpta_{sys1}$. We can see that the curve of $ra$ is almost close to the curve of $ra1$ when the chosen possibility is 31/32. When the chosen possibility is 1, $ra$ is equal to $ra1$, and their curve overlaps.

\begin{figure}
    \centering
    \includegraphics[width=3in]{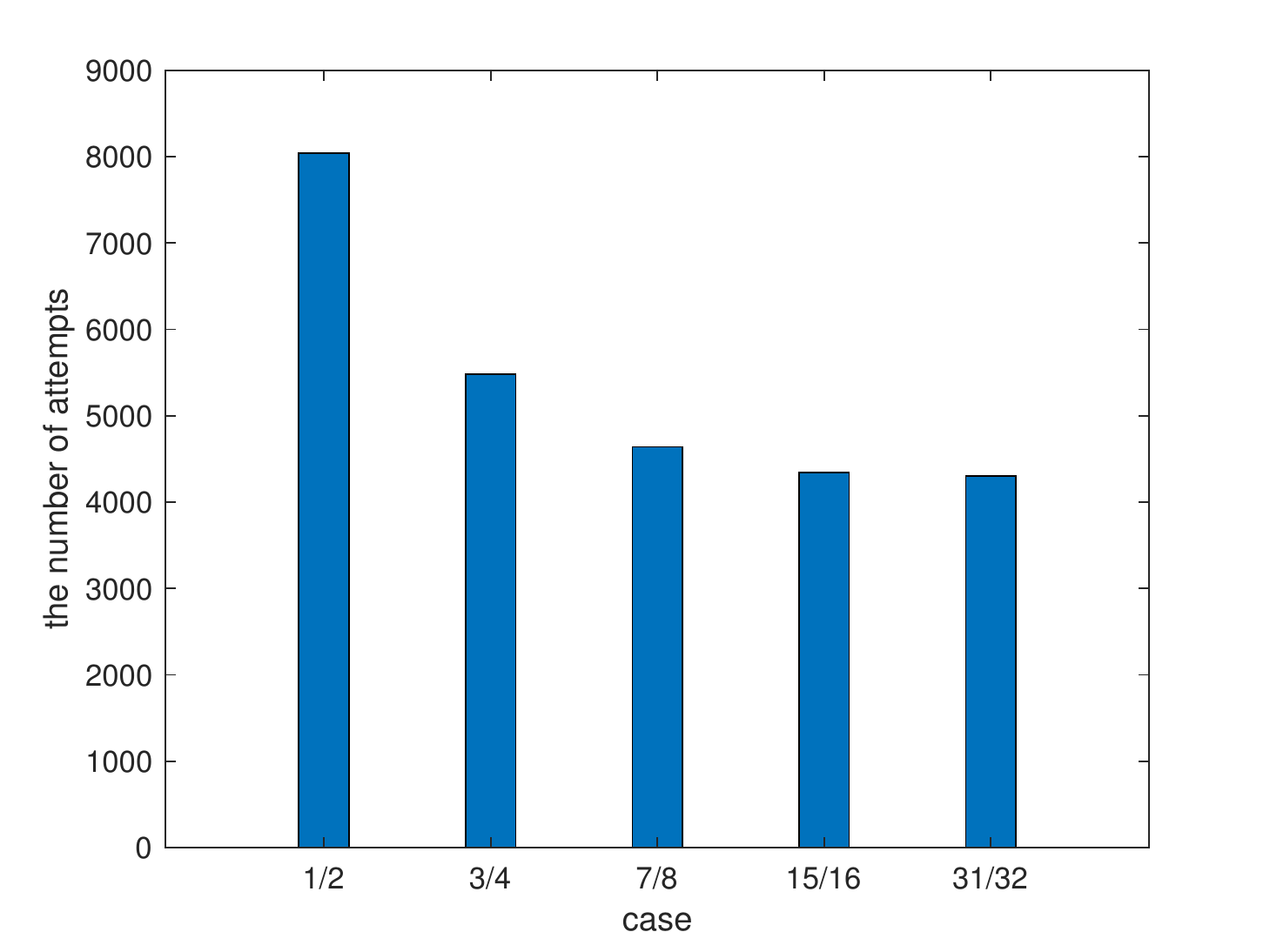}
    \caption{The average number of attempts for a successful attack with different choose possibility for the weaker systems (1/2, 3/4, 7/8,  and 15/16).}
    \label{averageNumberAttemps}
\end{figure}

From the average value of $ra$, we can also see that fewer attempts are required when the attacker knows which system is weaker. The average number of $ra$ is 8043.3 times when the chosen possibility is 1/2, and decreases to 4158.8 times when the chosen possibility is 1. This is shown in table \ref{t_average_attack_attempts} and Figure \ref{averageNumberAttemps}.

\begin{table}[!t]
    \caption{Average attack attempts}\label{t_average_attack_attempts}
     \centering
     \begin{tabular} {cccccc}
      \toprule
      Chosen possibility & 1/2 & 3/4 & 7/8 & 15/16 & 1\\
      \midrule
      Test & 8043.3 & 5479.8 & 4639.6 & 4339.5 & 4158.8\\
      Predicted of sys1 & 4096 & 4096 & 4096 & 4096 & 4096\\
      Predicted of sys2 & 16384 & 16384 & 16384 & 16384 & 16384\\      
      \bottomrule
    \end{tabular}
   \end{table}

Then, we get a conclusion that when the attacker knows which system is weaker, the whole system becomes weaker if the attacker chooses to attack the weaker system more frequently. Furthermore, when the attacker chooses the weaker system to attack with the chosen possibility 1, the whole system is as weak as the weaker system.

\subsection{Possibility to Run Two Consensus Algorithms}
In this section, we try to evaluate the resources occupied when a node runs the verification algorithms of two blockchains. It aims to show the possibility to use the CiFuV model. Two widely-used public blockchains, Bitcoin and Ethereum, are chosen in the verification.

We first show the minimum requirement to run those blockchain clients, and then we show the runtime resource occupation.

\subsubsection{Minimum Requirements of Hardware Resource to Run Bitcoin and Ethereum}
The minimum requirements to run a blockchain client can be seen as the static resource requirement. We get the minimum requirements from the Bitcoin official website and the Ethereum official website \footnote{https://ethereum.org/en/developers/docs/nodes-and-clients/}, which are shown in table \ref{t_min_requirement_bit_eth}. Row 1 shows the resource requirements of Bitcoin, and row 2 shows the resource requirements of Ethereum. Row 3 is the sum of those two clients, with the aim to have a rough assessment of the resource consumption when a node runs clients of both Bitcoin and Ethereum.

\begin{table*}[!t]
    \caption{Minimum requirements for the clients of Bitcoin and Ethereum}\label{t_min_requirement_bit_eth}
     \centering
     \begin{tabular}{p{12em}p{10em}p{16em}l}
      \toprule
      & CPU & Memory & Network\\
      \midrule
      Bitcoin client & 1 GHz & 1 GB & 5.5 GB/day\\
      Ethereum client & CPU with 2+ cores & 4 GB RAM minimum with an SSD, 8 GB+ if you have an HDD & 8 MBit/s\\
      Node with both clients (simply to sum Bitcoin and Ethereum) & CPU with 3+ cores with more than 1GHz & 5 GB RAM minimum with an SSD, 9 GB+ if you have an HDD & About 8 MB/s\\
      \bottomrule
    \end{tabular}
   \end{table*}

From table \ref{t_min_requirement_bit_eth}, we see the minimum hardware requirement for a node is not too high. The resources of a node with both clients can be supported by widely-used operating systems (OS), for example, CentOS. In fact, most operating systems have the capacity to support more CPUs and memory.

\subsubsection{Run-time Resource Consumption}
In this section, we try to show the resources occupied when the validation parts of Bitcoin and Ethereum are running on a node. It is called run-time resource consumption. As synchronization-keeping is proposed to synchronize information always instead of one-time synchronization, we show the result after the initial block synchronization.

The verification node is a virtual machine on an HP workstation (Z2G4), which is configured with i9 CPUs (16 logic processors @ 3.60 GHz), 64 GB memory, and two hard disks (256 GB SSD and 2 TB SATA). We assign 8 virtual processors and 8 GB of memory to this virtual machine. The operating system is Ubuntu 16.04.6 LTS (Xenial Xerus).

The verified resources include CPU consumption, memory consumption, and (network) bandwidth consumption. The data of the CPU and memory are from the output of the ‘top’ command (a Linux command to display Linux processes). The ‘\%CPU’ field of the output is for the CPU, which means the percent of one virtual processor. The '\%MEM' field is for the memory. The data of the bandwidth occupation is from the output of the ‘nethogs’ command (a Linux command to display bandwidth per process), which has two fields (one for the sending data, and another is for the receiving data).

The verification process is as follows. (1) We first start both clients, the client of Bitcoin (bitcoin-qt with version v0.21.0), and the clients of Ethereum (geth with version 1.10.1-stable). (2)A test script is used to collect the data every 1 minute, and a total of 480 samples are collected. In each sample, commands of ‘top’ and ‘nethogs’ are run. Then the corresponding data for the Bitcoin and Ethereum process are filtered and logged into separate files.

The consumption of CPU, memory, and network bandwidth is shown in Figure \ref{cpu_memory_network}.

From Figure \ref{cpu_memory_network}, we see that the CPU is less than 50\% of a virtual processor, and most samples (455 in all 480 samples) are less than 15\%. The memory consumption is less than 62\%, and most of the memory consumption (472 in all 480 samples) is between 50\% and 60\%. Most bandwidth used (462 in all 480 samples) is less than 100 KB/s.  Its average value is 46.31 KB/s. Only two times, the value of the bandwidth is bigger than 200 KB/s. 

The above results indicate that it is possible to run multiple blockchain clients on a single node. Nodes of one blockchain can get data from other blockchains and validate the data correctness in the CIFuV way.

\section{Related Work}
This section describes related work of the cross-chain interaction, which is in three aspects, atomicity of the cross-chain interaction, propagation methods of the cross-chain data, and related security issues.

Some work is to ensure atomicity during cross-chain interaction, which synchronizes states in different blockchains. The corresponding methods include the hash-locking method and the two-phase commitment. Hash-locking~\cite{deng2018research} locks a secret number in one blockchain, which is revealed when its owner (suppose user UA) tries to get another asset. In another blockchain, the paired exchanger can use this exposed number to get UA's assets. However, the number depends on users to pass it from one blockchain to another blockchain; while a user is not a publicly verifiable system. Even if the guest blockchain does not exist, fake information may be given to the host blockchain. In the two-phase commitment method~\cite{herlihy2018atomic}, related transactions are prepared in each blockchain, and the final commitment is done when both transactions appear in corresponding blockchains. This method helps to ensure that transactions are done synchronically. However, the verification of the correctness of transactions is not by the consensus of the guest blockchain, which may cause security downgrade issues. In summarization, there are security issues during the process to ensure the atomicity among blockchains currently. If the method is not fully trustful, the atomicity also has some challenges.

There are works for the information propagation among different blockchains, in which the information (transactions or blocks) propagates to another blockchain. Blockchain router~\cite{wang2017blockchain} is for this purpose. It connects different blockchains, collects related transactions, and forwards it to another blockchain. Work~\cite{kan2018multiple} propose to use one node from each blockchain; while this method only trusts one node of a blockchain, which may cause a security downgrade issue.

Another method to propagate the information is called the oracle technology, which introduces a third-party system. It is first in a centralized way~\cite{lo2020reliability}, and later its decentralized way is adopted. However, it does not validate the data by the consensus algorithm and the full data validation of the original blockchain, which may also have security downgrade issues.

There is also work which focuses on the security issues of the blockchain\cite{li2020survey}~\cite{lin2017survey}\cite{ moubarak2018blockchain}; while they focus on one blockchain and does not analyze the security issues of the cross-chain interaction. Those works can mainly be divided into three types, the security issues of the P2P networks, the security issues of the consensus algorithm and the security issues of the smart contracts. 

Related work for the security issues of the P2P networks. Work~\cite{marcus2018low} analyzes the eclipse attack that monopoly the connection of the target node and makes the victim node not able to get the information from other nodes of the blockchain. The data propagation time over the P2P network plays a significant role on a permissionless blockchain, and work~\cite{rahmadika2020dilemma} focuses on the propagation issues and analyzes the related possible attack.

Related work for the security issues of the consensus algorithm. Work~\cite{bach2018comparative} compares several aspects of the consensus algorithm including their security. For the PoW consensus, if the hashing power of certain node(s) exceeds 51\% of that of the whole blockchain, it can rebranch that blockchain to achieve double-spending. Work~\cite{sayeed2019assessing} analyzes the 51\% attack and the most advanced protection methods. Self-mining is one of the attacks, in which malicious nodes do not expose their newly-mined blocks immediately to achieve an extra reward for those nodes. Work~\cite{chicarino2020detection} proposes method to detect this behavior. Work~\cite{mirkin2020bdos} analyzes the blockchain DoS (BDoS). BDoS discourages the participation of other minders by exploiting the reward mechanism, and it only needs significantly fewer resources than 51\% (such as 21\%).

Related work for the security of the smart contract. (1)Some works focus on improving the smart contract programming language. Work~\cite{pettersson2016safer} proposes a type-driven development language, which adopts dependent and polymorphic types to make smart contracts safer. Work~\cite{schrans2018writing} proposes a contract-oriented programming language, which is type-safe and capabilities-secure, with the aim to program robust smart contracts. Work~\cite{coblenz2017obsidian} proposes a user-centered approach language, which facilitates the programming of smart contracts. Work~\cite{lin2017survey} proposes a typed combinator-based language, with the aim to eliminate loops and recursion in the smart contracts. (2) Some work aims to formally verify the smart contract to find potential bugs in advance. Work~\cite{hildenbrandt2018kevm} focuses on the EVM byte code of the smart contract, which aims to give a formal specification for further analysis. Work~\cite{hirai2017defining} proposes the interactive theorem provers for the smart contract. To analyze the runtime safety and functional correctness of smart contracts, work~\cite{bhargavan2016formal} proposes a framework that translates a smart contract into a functional language to do the verification.

\section{Conclusion}
In this paper, we focus on the security issues of the cross-chain interaction. We first propose the data interaction model to analyze the process of the system interaction and point out two possible security issues. One is the security downgrade issue — a more secure system may lose its security by a less secure method or a less secure system. Another one is the blockchain invisible issue — the cross-chain interaction goes on in a blockchain even if its paired system does not exist. Then, to overcome the above two issues, we propose the fully verified method as the cross-chain interaction model. This method adapts the consensus algorithm of the guest blockchain to verify the validation of the data on the guest blockchain. At last, we evaluate the data interaction model and demonstrate the possible security issues. Meanwhile, we also evaluate the possibility that one node to runs two blockchain clients, aiming to show that it is possible to verify another blockchain in the host blockchain.

\bibliographystyle{unsrt}

\bibliography{mar}

\end{document}